\documentclass[twocolumn,showpacs,preprintnumbers,amsmath,amssymb]{revtex4}

\usepackage{graphicx}
\usepackage{dcolumn}
\usepackage{bm}

\begin{document}

\title{A directed network model for World-Wide Web}
\author{Jian-Guo Liu}
\author{Yan-Zhong Dang}
\author{Zhong-Tuo Wang}
\affiliation{ Institute of System Engineering, Dalian University of
Technology, Dalian Liaoning, 116023, PR China
}%
\author{Tao Zhou}
\email{zhutou@ustc.edu}
\affiliation{Department of Modern Physics,
University of Science and Technology of China, Hefei Anhui, 230026,
PR China}
\date{\today}

\begin{abstract}
In this paper, a directed network model for world-wide web is
presented. The out-degree of the added nodes are supposed to be
scale-free and its mean value is $m$. This model exhibits
small-world effect, which means the corresponding networks are of
very short average distance and highly large clustering coefficient.
More interesting, the in-degree distribution obeys the power-law
form with the exponent $\gamma=2+1/m$, depending on the average
out-degree. This finding is supported by the empirical data, which
has not been emphasized by the previous studies on directed
networks.
\end{abstract}


\maketitle

\section{Introduction}
The last few years have burst a tremendous activity devoted to the
characterization and understanding of complex
network\cite{S2001,AB02,DM02,N2003}. Researchers described many
real-world systems as complex networks with nodes representing
individuals or organizations and edges mimicking the interaction
among them. Commonly cited examples include technological networks,
information networks, social networks and biological networks
\cite{N2003}. The results of many experiments and statistical
analysis indicate that the networks in various fields have some
common characteristics. They have small average distances like
random graphs, large clustering coefficients like regular networks,
and power-law degree distributions. The above characters are called
the small-world effect\cite{WS98} and scale-free
property\cite{BA99}.

Motivated by the empirical studies on various real-life networks,
some novel network models were proposed recently. The first
successful attempt to generate networks with high clustering
coefficient and small average distance is that of Watts and Strogatz
(WS model) \cite{WS98}. The WS model starts with a ring lattice with
$N$ nodes wherein every node is connected to its first $2m$
neighbors. The small-world effect emerges by randomly rewiring each
edge of the lattice with probability $p$ such that self-connections
and duplicate edges are excluded. The rewiring edges are called
long-range edges which connect nodes that otherwise may be part of
different neighborhoods. Recently, some authors have demonstrated
that the small-world effect can also be produced by using
deterministic methods\cite{CS,CF,ZWJX}.

Another significant model capturing the scale-free property is
proposed by Barab\'{a}si and Albert (BA network) \cite{BA99,BAJ99}.
Two special features, i.e., the growth and preferential attachment,
is investigated in the BA networks for the free scaling of the
Internet, WWW and scientific co-authorship networks, etc. These
points to the fact that many real-world networks continuously grow
by the way that new nodes added to the network, and would like to
connect to the existing nodes with large number of neighbors.

While the BA model captures the basic mechanism which is responsible
for the power-law distribution, it is still a minimal model with
several limitations: it only predicts a fixed exponent in a
power-law degree distribution, and the clustering coefficients of BA
networks is very small and decrease with the increasing of network
size, following approximately $C\sim ln^2N/N$\cite{Klemm2002}. To
further understand various microscopic evolution mechanisms and
overcome the BA model's discrepancies, there have been several
promising attempts. For example, the aging effect on nodes' charms
leads the studies on the {\bf aging
models}\cite{Klemm2002,Amaral2000,SND2000,Jiang2005}, the
geometrical effect on the appearance probability of edges leads the
studies on the {\bf networks in Euclidean space}\cite{ES1,ES2,ES3},
and the self-similar effect on the existence of hierarchical
structures leads the studies on the {\bf hierarchical
models}\cite{RB,DGM,AHAS,ZYW,GU,ZZZ}.

One of the extensively studied networks is the World-Wide
Web\cite{WWW1,WWW2,WWW3,WWW4}, which can be treated as a directed
network having power-law distributions for both in-degree and
out-degree. In addition, it is a small-world networks. Since the
knowledge of the evolution mechanism is very important for the
better understanding of the dynamics built upon WWW, many
theoretical models have been constructed
previously\cite{BE2001,T2001,T2002}. However, these models haven't
considered the relationship between the in-degree distribution and
the out-degree distribution.

\begin{figure}
{\includegraphics[width=0.45\textwidth]{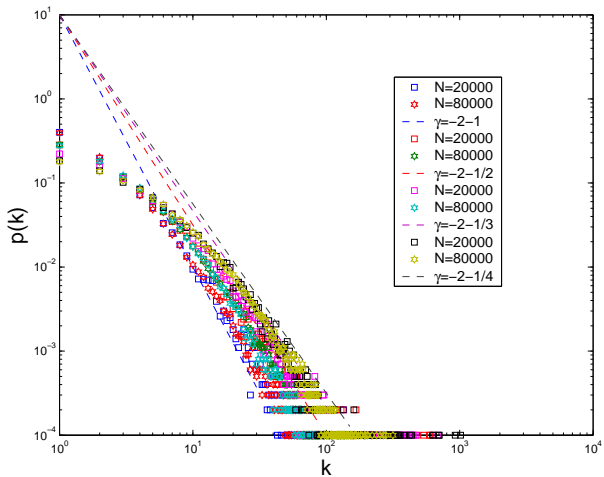}}
\caption{Degree distributions for different $N$ and $m$. In this
figure, $p(k)$ denotes the probability that a randomly selected node
is of in-degree $k$. When $m=1$, the power-law exponent $\gamma$ of
the density functions
 are $\gamma_{20000}=2.95\pm 0.06$ and $\gamma_{80000}=2.97\pm
 0.04$. When $m=2$, $\gamma_{20000}=2.46\pm 0.07$ and $\gamma_{80000}=2.47\pm
 0.03$. When $m=3$, $\gamma_{20000}=2.29\pm 0.08$ and $\gamma_{80000}=2.31\pm
 0.03$. When $m=4$, $\gamma_{20000}=2.21\pm 0.07$ and $\gamma_{80000}=2.23\pm
 0.03$. The four dash lines of $m=1,2,3,4$ have slope -3, -2-1/2,
 -2-1/3 and -2-1/4 for comparison, respectively.}
\end{figure}

In this paper, we propose a directed network model for the
World-Wide Web. This model displays both scale-free and small-world
properties, and its power-law exponent of out-degree distribution is
determined by the average in-degree. Comparisons among the empirical
data, analytic results and simulation results strongly suggest the
present model a valid one. The rest of this paper is organized as
follows: In section 2, the present model is introduced. In section
3, the analyzes and simulations on network properties are shown,
including the degree distribution, the average distance, and the
clustering coefficient. Finally, in section 4, the main conclusion
is drawn.

\section{The model}
Our model starts with a connected graph of $N_0$ nodes and $m_0$
edges. At each time step $i$, a new node $v_i$ is added and $2e_i$
existing nodes are chosen to be its neighbors. The choosing
procedure involves two processes: preferential attachment\cite{BA99}
and neighboring attachment\cite{HK2002}. Firstly, in the
preferential attachment process, $e_i$ nodes, denoted by the set
$Q_i$, are selected with probability proportional to their
in-degrees. And then, in the neighboring attachment process, for
each node $x\in Q_i$, one of its neighbors is randomly selected to
connect to $v_i$. Combine these two processes, there are in total
$2e_i$ nodes having been chosen as the new one's neighbors. In the
whole evolution processes, the self-connections and duplicute edges
are excluded.

It should be emphasized that, since the out-degree of the WWW
network is not fixed but approximately obeying a power law, the
number of newly added edges during one time step, $2e$, is not a
constant but a random number also obeying a power-law. And the
average out-degree $m$ is fixed, which significantly affects the
in-degree distribution exponent, average distance and clustering
coefficient of the whole network.

\section{The Statistical Characteristics}
In this section, the scale-free small-world characteristics of the
present model are shown.

\subsection{The Scale-free Property}

The probability that a newly appearing node connects to a previous
node is simply proportional to the in-degree $k$ of the old vertex.
Suppose the newly added node's attraction is $A$, then the
probability of attachment to the old vertices should be proportional
to $k+A$, where $A$ is a constant and we set $A=1$ for
simplicity\cite{DMS2000}. The probability that a new edge attaches
to any of the vertices with degree $k$ is
\begin{equation}\label{F3.1}
\frac{(k+1)p_k}{\sum_k (k+1)p_k}=\frac{(k+1)p_k}{m+1}.
\end{equation}

The mean out-degree of the newly added node is simply $m$, hence the
mean number of new edges to vertices with current in-degree $k$ is
$(k+1)p_km/(m+1)$. Denote $p_{k,n}$ the value of $p_k$ when the
network size is $n$, then the change of $np_k$ is
\begin{equation}
\left\{
    \begin{array}{ll}
      (n+1)p_{k,n+1}-np_{k,n}=\frac{m[kp_{k-1,n}-(k+1)p_{k,n}]}{m+1}
      & k\geq 1 \\[5pt]
      (n+1)p_{0,n+1}-np_{0,n}=1-p_{0,n}\frac{m}{m+1} & k=0 \\
    \end{array}
\right.
\end{equation}
The stationary condition $p_{k,n+1}=p_{k,n}=p_k$ yields
\begin{equation}\label{F3.4}
p_k=\left\{
        \begin{array}{lc}
        [kp_{k-1}-(k+1)p_k]m/(m+1), & \  k\geq 1; \\[5pt]
        1-p_0 m/(m+1),              & \  k=0.
        \end{array}
   \right.
\end{equation}
Rearranging, one gets
\begin{equation}\label{F3.4}
p_k=\left\{
        \begin{array}{lc}
         \frac{k}{k+2+1/m}p_{k-1}, & \  k\geq 1; \\[5pt]
        (m+1)/(2m+1),              & \  k=0.
        \end{array}
   \right.
\end{equation}
This yields
\begin{equation}\label{F3.5}
\begin{array}{rcl}
p_k & = & \frac{k(k-1)\cdots 1}{(k+2+1/m)\cdots (3+1/m)}p_0\\[5pt]
    & = & (1+1/m)B(k+1,2+1/m),
\end{array}
\end{equation}
where $B(a,b)=\Gamma(a)\Gamma(b)/\Gamma(a+b)$ is Legendre's beta
function, which goes asymptotically as $a^{-b}$ for large $a$ and
fix $b$, hence
\begin{equation}\label{F3.6}
p_k\sim k^{-(2+1/m)}.
\end{equation}
This leads to $p_k\sim k^{-\gamma_i}$ with $\gamma_i=(2+1/m)$ for
large $N$, where $\gamma_i$ is the exponent of the in-degree degree
distribution.

In Fig. 1, the degree distributions for $m=1,2,3,4$ are shown. The
simulation results agree with the analytic one very well and
indicate that the exponents of the degree distribution have no
relationship to the network size $N$.

\begin{figure}[ht]
    \begin{center}
       {\includegraphics[width=0.45\textwidth]{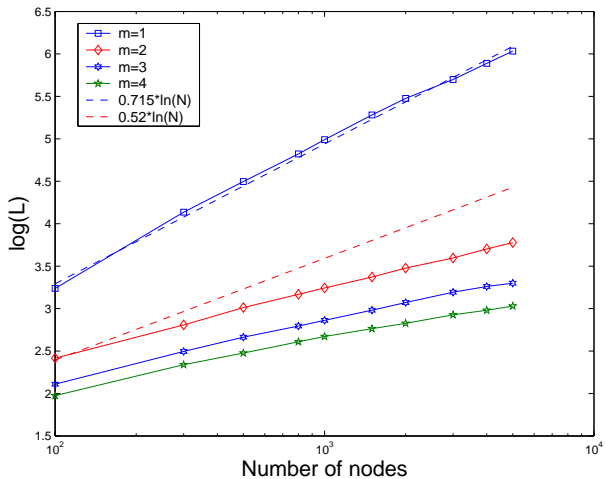}}
 \caption{\small The average distance $L$ vs network size $N$ of the undirected version of the present model. One can see that $L$
 increases very slowly as $N$ increases. The main plot exhibits the curve where
 $L$ is considered as
 a function of ${\rm ln}N$, which is well fitted by a straight line.
 When $m=1$, the curve is above the fitting line when $N\leq 3000$
 and under the line when $N\geq 4000$. When $m=2,3,4$, the curve
 is under the line when $N\geq 200$, which indicates that the
 increasing tendency of $L$ is approximately to ${\rm ln}N$, and
 in fact a little slower than ${\rm ln}N$.}
 \end{center}
\end{figure}

One of the significant empirical results on the in- and out-degree
distributions is reported by Albert, Jeong and Barab$\acute{a}$si
\cite{AJB1999}. In this paper the crawl from Altavista was used. The
appearance of the WWW from the point of view of Altavista is as
following \cite{DM02}:

\begin{itemize}
\item In May 1999 the Web consisted of 203$\times 10^6$ vertices
and 1466$\times 10^6$ hyperlinks. The average in- and out-degree
were $\overline{k}_{in}=\overline{k}_{out}=7.22$.

\item In October 1999 there were already 271$\times 10^6$ vertices
and 2130$\times 10^6$ hyperlinks. The average in- and out-degree
were $\overline{k}_{in}=\overline{k}_{out}=7.85$.
\end{itemize}
The distributions were found to be of a power-law form with exponent
$\gamma_i=2.1$ and $\gamma_o=2.7$, where $\gamma_o$ is the exponent
of the out-degree degree distribution. When
$\overline{k}_{out}=7.22$ and $7.85$, one can obtained from
$\gamma_i=2+1/m$ that $\gamma_i=2.138$ and $2.127$ respectively,
which is very close to 2.1, thus give a strong support to the
validity of the present model.

\subsection{The Average Distance}
The average distance plays a significant role in measuring the
transmission delay, thus is one of the most important parameters to
measure the efficiency of communication network. Since the original
conception of small-world effect is defined based on undirected
networks, hereinafter we only consider the undirected version of our
model, that is, the directed edge $E_{ij}$ from node $i$ to $j$ is
considered to be an bidirectional edge between node $i$ and $j$.
When the node is added to the network, each node of the network
according to the time is marked. Denote $d(i,j)$ the distance
between nodes $i$ and $j$, the average distance with network size
$N$ is defined as
\begin{equation}\label{F3.1}
L(N)=\frac{2\sigma(N)}{N(N-1)},
\end{equation}
where the total distance is:
\begin{equation}\label{F33.1}
\sigma(N)=\sum_{1\leq i<j\leq N}d(i,j).
\end{equation}
Clearly, the distance between the existing nodes will not increase
with the network size $N$, thus we have
\begin{equation}\label{F3.2}
\sigma(N+1)\leq \sigma(N)+\sum_{i=1}^N d(i,N+1).
\end{equation}
Denote $y=\{y_1, y_2, \cdots, y_l\}$ as the node set that the
$(N+1)${\rm th} node have connected. The distance $d(i,N+1)$ can be
expressed as following
\begin{equation}\label{F3.2}
d(i,N+1)=\min \{d(i,y_j)|j=1,2,\cdots,l\}+1.
\end{equation}
Combining the results above, we have
\begin{equation}\label{F33.2}
\sigma(N+1)\leq \sigma(N)+(N-l)+\sum_{\Lambda}D(i,y),
\end{equation}
where $\Lambda=\{1, 2, \cdots, N\}-\{y_1, y_2, \cdots, y_l\}$ is a
node set with cardinality $N-l$. Consider the set $y$ as a single
node, then the sum $\sum_{i=\Lambda}d(i,y)$ can be treated as the
distance from all the nodes in $\Lambda$ to $y$, thus the sum
$\sum_{i=\Lambda}d(i,y)$ can be expressed approximately in terms of
$L(N-l)$
\begin{equation}\label{F3.5}
\sum_{i=\Lambda}d(i,y)\approx (N-l)L(N-l).
\end{equation}
Because the average distance $L(N)$ increases monotonously with $N$,
this yields
\begin{equation}\label{F3.6}
(N-l)L(N-l)=(N-l)\frac{2\sigma(N-l)}{(N-l)(N-l-1)}<\frac{2\sigma(N)}{N-l-1}.
\end{equation}
Then we can obtain the inequality
\begin{equation}\label{F3.7}
\sigma(N+1)<\sigma(N)+(N-l)+\frac{2\sigma(N)}{N-l-1}.
\end{equation}
Enlarge $\sigma(N)$, then the upper bound of the increasing tendency
of $\sigma(N)$ will be obtained by the following equation.
\begin{equation}\label{F3.8}
\frac{d\sigma(N)}{dN}=N-l+\frac{2\sigma(N)}{N-l-1}.
\end{equation}
This leads to the following solution:
\begin{equation}\label{F3.9}
\sigma(N)=(N-l-1)^2{\rm log}(N-l-1)-(N-l-1)+C_1(N-l-1).
\end{equation}

\begin{figure}[ht]
    \begin{center}
       {\includegraphics[width=0.45\textwidth]{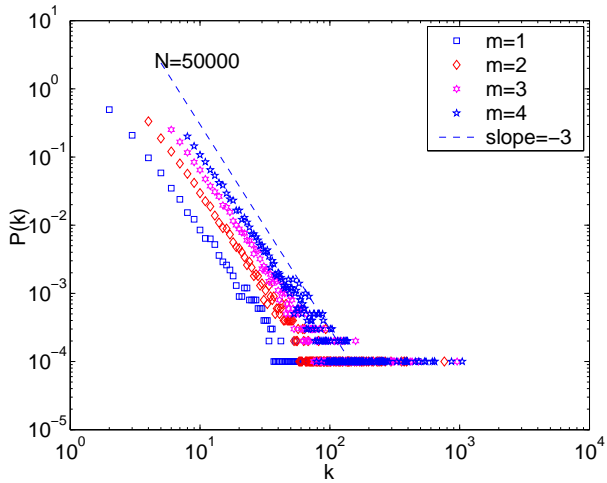}}
 \caption{Degree distribution of the undirected versions of the present model. At each time
 step, the new node selects $m=1,2,3,4$ edges to connected, respectively.
 When $m=1,2,3,4$, the power-law exponent $\gamma$ of the density functions
 are $\gamma_{1,80000}=2.95\pm 0.06$, $\gamma_{2,80000}=2.97\pm
 0.05$, $\gamma_{3, 80000}=2.96\pm 0.07$ and $\gamma_{4,80000}=2.96\pm
 0.06$, respectively. The dash line have slope -3.0 for comparison.}
 \end{center}
\end{figure}

From Eq.(\ref{F3.1}), we have that $\sigma(N)\sim N^2L(N)$, thus
$L(N)\sim {\rm ln}N$. Since Eq.(\ref{F3.7}) is an inequality, the
precise increasing tendency of the average distance $L(N)$ may be a
little slower than ${\rm ln}N$. The simulation results are reported
in figure 2.

\subsection{The Clustering Coefficient}

The clustering coefficient is defined as
$C=\sum_{i=1}^N\frac{C_i}{N}$, where
\begin{equation}\label{F4.1}
C_i=\frac{2E(i)}{k_i(k_i-1)}
\end{equation}
is the local clustering coefficient of node $i$, and $E(i)$ is the
number of edges among the neighboring set of node $i$.
Approximately, when the node $i$ is added to the network, it is of
degree 2$e_i$ and $E(i)\approx e_i$ if the network is sparse enough.
And under the sparse case, if a new node is added as $i$'s neighbor,
$E(i)$ will increase by 1. Therefore, in terms of $k_i$ the
expression of $E(i)$ can be written as following:
\begin{equation}\label{F4.2}
E(i)=e_i+(k_i-2e_i)=k_i-e_i.
\end{equation}
Hence, we have
\begin{equation}\label{F4.3}
C_i=\frac{2(k_i-e_i)}{k_i(k_i-1)}.
\end{equation}
This expression indicates that the local clustering scales as
$C(k)\sim k^{-1}$. It is interesting that a similar scaling has been
observed in pseudofractal web \cite{DGM} and several real-life
networks \cite{RB}. In figure 4, we report the simulation result
about the relationship between $C(k)$ and $k$, which is in good
accordance with both the analytic and empirical data \cite{RB}.

Consequently, we have
\begin{equation}\label{F4.41}
C=\frac{2}{N}\sum^N_{i=1}\frac{k_i-e_i}{k_i(k_i-1)}=\frac{2}{N}\sum^N_{i=1}
\frac{k_{\rm in}}{k_i(k_i-1)},
\end{equation}
where $k_{\rm in}$ denotes the in-degree of the $i$th node. Because
the average out-degree is $m$, one can replace the out-degree of
each node by $m$. From Fig. 3, one can get that the degree
distribution of the undirected network is $p(k)\sim k^{-3}$, where
$k=k_{\rm min}, k_{\rm min}+1, \cdots, k_{\rm max}$. As an example,
the clustering coefficient $C$ when $m=1$ can be rewritten as
\begin{equation}\label{F4.4}
C=\frac{2}{N}\sum^N_{i=1}\frac{1}{k_i}.
\end{equation}
Since the degree distribution is $p(k)=c_1 k^{-3}$, where
$k=2,3,\cdots, k_{\rm max}$. The clustering coefficient $C$ can be
rewritten as
\begin{equation}
C=\sum_{k=2}^{k_{\rm max
}}\frac{2}{N}\frac{Np(k)}{k}=2c_1\sum_{k=2}^{k_{\rm max}} k^{-4}.
\end{equation}
For sufficient large $N$, $k_{max}\gg 2$. The parameter $c_1$
satisfies the normalization equation
\begin{equation}\label{F4.8}
\sum_{k=2}^{k_{\rm max}}p(k)dk=1.
\end{equation}

\begin{figure}[ht]
    \begin{center}
       {\includegraphics[width=0.45\textwidth]{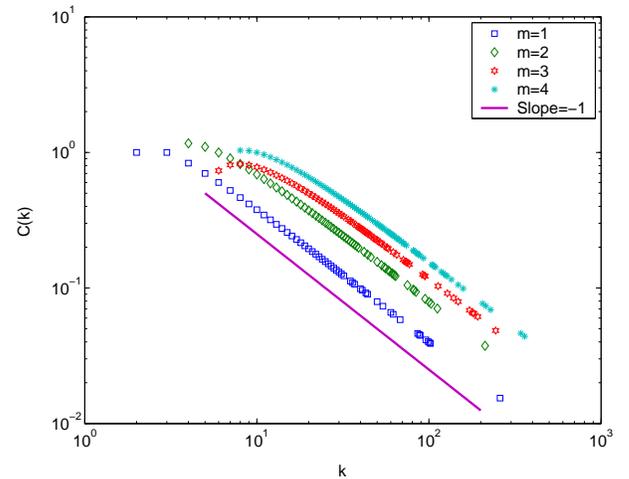}}
 \caption{Dependence between the clustering coefficient and the
 degree $k$ when $N=2000$. One can see that the clustering coefficient and
 the degree $k$ follow the reciprocal law.
 }\label{Fig5}
 \end{center}
\end{figure}

\begin{figure}[ht]
    \begin{center}
       {\includegraphics[width=0.45\textwidth]{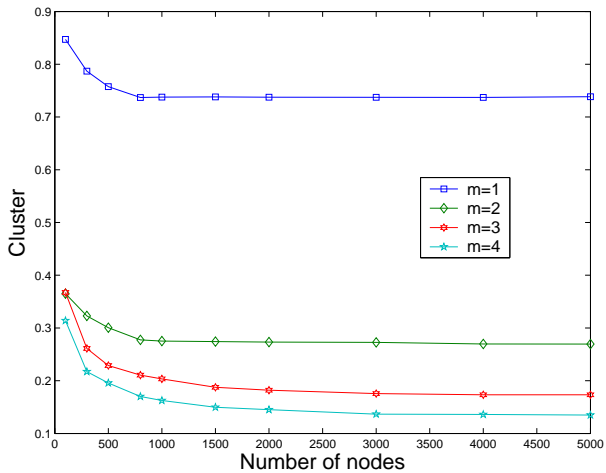}}
 \caption{The clustering coefficient vs the network size $N$ to different $m$ of the undirected versions of the present model.
 In this figure, when $m=1,2,3,4$, one can find that the clustering coefficient of
 the network is almost a constant 0.74, 0.28, 0.18 and 0.14,
 respectively. This indicates that the average clustering
 coefficient is relevant to the average out-degree $m$.
 }
 \end{center}
\end{figure}

It can be obtained that $c_1=4.9491$ and $C=2\times
4.9491\times\sum_{k=2}^{k_{\rm max}}k^{-4}=0.8149$. The
demonstration exhibits that most real-life networks have large
clustering coefficients no matter how many nodes they have. From
Fig. 5, one can get that as the average out-degree increases, the
clustering coefficient decreases dramatically, which indicates that
the clustering coefficient $C$ is relevant to the average out-degree
$m$.

\section{Conclusion and Discussion}
In summary, we have constructed a directed network model for
World-Wide Web. The presented networks are both of very large
clustering coefficient and very small average distance. We argue
that the degree distribution of many real-life directed networks may
be fitted appropriately by two power-law distributions, i.e., in-
and out-degree power-law distributions, such as the citation
network, Internet network and World-Wide Web. Both the analytic and
numerical studies indicate the exponent of the in-degree
distribution of the presented networks can be well fitted by
$2+1/m$, which has been observed in the empirical data. Although
this model is simple and rough, it offers a good starting point to
explain the existing empirical data and the relationship between the
in- and out-degree distribution exponents.

\subsection*{Acknowledgment}
The authors are grateful to Dr. Qiang Guo for her valuable comments
and suggestions, which have led to a better presentation of this
paper. This work has been supported by the National Science
Foundation of China under Grant Nos. 70431001 and 70271046.

\end{document}